\documentclass[aip,jcp,reprint,floatfix]{revtex4-1}

\usepackage{amssymb}
\usepackage{amsmath}
\usepackage{amsfonts}
\usepackage{graphicx}

\usepackage{longtable}

\newcommand{\ls}{\lesssim}

\newcommand{\lbr}{\left(}
\newcommand{\rbr}{\right)}

\renewcommand{\vec}[1]{\boldsymbol{#1}}

\newcommand{\beq}{\begin{eqnarray}}
\newcommand{\eeq}{\end{eqnarray}}
\renewcommand{\d}{{\text{d}}}

\newcommand{\rcite}[1]{Ref. \onlinecite{#1}}

\newcommand{\Hxc}{\text{Hxc}}

\newcommand{\pr}{^{\prime}}

\newcommand{\vr}{\vec{r}}

\newcommand{\vrp}{\vec{r}\pr}

\newcommand{\Ext}{{\text{Ext}}}

\newcommand{\LEXX}{{\text{LEXX}}}

\renewcommand{\k}{\kappa}

\usepackage{color,xcolor}
\usepackage[normalem]{ulem}
\begin{document}
\title{How polarizabilities and $C_6$ coefficients %
  actually vary with atomic volume}
\author{Tim Gould}\affiliation{Qld Micro- and Nanotechnology Centre, %
Griffith University, Nathan, Qld 4111, Australia}
\begin{abstract}
  In this work we investigate how atomic $C_6$ coefficients and
  static dipole polarizabilities $\alpha$ scale with effective volume.
  We show, using confined atoms covering rows 1-5 of the periodic table,
  that $C_6/C_6^R\approx (V/V^R)^{p_Z}$
  and $\alpha/\alpha^R\approx (V/V^R)^{p'_Z}$
  (for volume $V=\int d\vr \frac{4\pi}{3}r^3 n(\vr)$)
  where $C_6^R$, $\alpha^R$ and $V^R$ are
  the reference values and effective volume of the free atom.
  The scaling exponents $p_Z$ and $p'_Z$ vary substantially
  as a function of element number $Z=N$, in contrast to the
  standard ``rule of thumb'' that $p_Z=2$ and $p'_Z=1$.
  Remarkably, We find that the polarizability and $C_6$ exponents
  $p'$ and $p$ are related by
  $p'\approx p-0.615$ rather than the expected $p'\approx p/2$.
  Results are largely independent of the form of the confining
  potential (harmonic, cubic and quartic potentials are considered)
  and kernel approximation, justifying this analysis.
\end{abstract}
\maketitle

\section{Introduction}

In recent years a great deal of attention has been paid to 
embedding theories, especially for methods that introduce
dispersion forces to \emph{ab initio} calculations.
At least two popular methods, Becke and Johnson's\cite{Becke2005-XDM}
exchange dipole model (XDM),
and Tkatchenko and Scheffler's\cite{Tkatchenko2009} approach (TS)
and its derivatives\cite{Tkatchenko2012,DiStasio2012,Ambrosetti2016},
are based on a model of dispersion forces that include atom-wise
contributions based on reference pro-atom $C_6$ coefficients
and/or static polarizabilities $\alpha(0)$.
These pro-atom properties are rescaled
by the square of the normalised effective volume of the atoms i.e.
that
\begin{align}
  \frac{C_6(V)}{C_6(V^R)}=&\lbr \frac{V}{V^R} \rbr^2
  \label{eqn:Scale0}
\end{align}
to account for the effects of confinment.
Here $C_6^R\equiv C_6(V^R)$ is the reference atomic coefficient,
\begin{align}
  V^R=&\int d\vr n^R(\vr) \frac{4\pi r^3}{3}
  \label{eqn:V}
\end{align}
is the reference atomic volume,
and $V$ is an effective volume of the atom in the
molecule or bulk. A similar power law is employed for the dipole
polarizability $\alpha$, but with an exponent of one.

This rescaling is included to account for the confining effect of
the other electrons and nucleii in a molecule.
However, despite wide-ranging successes, the scaling assumption that
goes into XDM, TS and successor approximations is based on a somewhat
limited range of data. The validity of \eqref{eqn:Scale0}
has implications for any theory that relies on re-scaled, 
pre-calculated atomic polarizability data, whether for
van der Waals forces or for more general force field models.
We will show, in this work, that this ``rule of thumb''
assumption is not correct for real atoms. Consequently,
care should be taken when employing it.

\section{Theory}

The origins of the ``rule of thumb'' come from a number of
different directions.
Equation~\eqref{eqn:Scale0} can be show analytically
for hard-sphere models and was was extended semi-analytically
to the context of atoms by Dmitrieva and Plindov\cite{Dmitrieva1983}.
Later work on atoms in molecules
by Brinck, Murray and Politzer\cite{Brinck1993} showed that the
relationship held also in molecular cases. However, there results
were restricted almost entirely to first row elements, with only
a few examples of larger atoms.

In more recent work, Politzer, Jin and Murray\cite{Politzer2002}
explored the proportionality of free atomic polarizability
properties with different definitions of atomic volume.
Even more recently,
Kannemann and Becke\cite{Kannemann2012} used the XDM model
to study correlations between free atom polarizabilities and
atomic volumes, highlighting basis set and density functional
approximation variations.
Blair and Thakkar\cite{Blair2014} explored different relationships
in a large database of molecules, showing that the inverse square
average of momentum might be a useful quantity when looking for
relationships between polarizability and effective volume.
While both these works are very interesting and comprehensive in
their analysis, neither addresses directly how the volume
of atoms embedded in a molecule changes polarizability.

In a slightly different context, the study of atomic properties
in confined potentials has been a long-standing topic of
interest (see e.g. the recent collections in
Refs~\onlinecite{Cruz2009-1} and \onlinecite{Cruz2009-2}).
Most of these studies are restricted to one and two-electron
systems with limited attention on open shell systems, however.

In this work, rather than looking for relationships between
isolated atomic and molecular properties which can be applied
to atoms in molecules, we will adopt the direct approach
and explore the relationship between the specific definition
of volume given above in \eqref{eqn:V} and static
dipole polarizabilities and $C_6$ coefficients
in \emph{confined} atoms
designed to mimic embedded atoms.
We will show that both $\alpha$ and $C_6$ relate to volume
via power laws, but that the exponent varies signficantly
for different atoms.
We will study all atoms in rows 1-5 of the periodic table,
avoiding any bias towards closed shells or typically
``organic'' elements.

Specifically, we will numerically study model electronic systems
with an external potential
\begin{align}
  v_{\Ext}(r)=&\frac{-Z}{r} + \frac{r^{\k}}{r_c^{\k+1}}
\end{align}
comprising a standard atomic potential $-Z/r=-N/r$ and an
additional, confining harmonic ($\k=2$), cubic ($\k=3$) or
quartic ($\k=4$) potential $r^{\k}/r_c^{\k+1}$ that mimics the effect of
the neighbouring atoms in the molecule as controlled by $r_c$
, where $r_c$ is approximately the distance at which the
total external potential is zero%
\footnote{We work in atomic units here and throughtout this work. Thus
volumes and polarizabilities have dimensions of $a_0^3$ (where
$a_0$ is the Bohr radius), $C_6$ coefficients have units
of Ha$a_0^6$, and effective frequencies have units of Ha$^{-1}$.}.
We will show that scaling is essentially independent of $\k$,
suggesting a certain universality.
We will then use this study to derive relationships between the
properties of the unscaled coefficients and polarizabilities and
their scaling exponents.

\subsection{Methodology}

All calculations for this work are carried out using all-electron,
linear-response, time-dependent density functional theory (tdDFT).
In the atomic systems tested, linear-response tdDFT offers a
significant speed advantage over conventional
high-level many-electron methods,
while offering a level of acuracy that exceeds conventional
groundstate techniques (see e.g. the discussion in
Refs.~\onlinecite{Eshuis2012,Ren2012,Dobson2012-JPCM}).
It thus offers the ability to test large numbers of atomic
systems relatively quickly.
The tdDFT approach has previously been employed in this context
by Chu and Dalgarno\cite{Chu2004} and
by Ludlow and Lee\cite{Ludlow2015}. Even more recently,
Gould and Bu\v{c}ko\cite{Gould2016-C6} used
tdDFT to successfully evaluate polarizabilities and
$C_6$ coefficients for rows~1-6 of the periodic table.

This work follows closely the calculations of \rcite{Gould2016-C6}.
However, unlike that work we are unable to use reference
data to reintroduce relativistic effects, and thus
do not consider row~6.
These calculations employ both PGG and RXH approximate
kernels (discussed later),
which introduces an error to all polarizabilities.
As Chu and Dalgarno\cite{Chu2004} note, the
kernel approximations tend to introduce consistent errors
for different species, an observation backed by
\rcite{Gould2016-C6}. We thus assume (and will later show)
that, while the kernels may not reproduce
\emph{quantitative polarizabilities}, they can certainly
reproduce \emph{quantitative trends in polarizabilities}.

\subsection{Technical details}

Since self-interaction and static correlation errors contribute to
the dipole response of even large atoms, we employ the
linear exact exchange (LEXX) functional\cite{Gould2013-LEXX},
based on ensemble DFT\cite{Valone1980} in both the groundstate
and linear response calculations.
LEXX extends the good self-interaction physics of the
exact exchange (EXX) functional\cite{Kummel2008}
to open-shell systems while formally maintaining numerically
efficient spherical symmetry, giving access to all atoms in
the tested rows 1-5.
It thus allows us to evaluate asymptotically accurate Kohn-Sham (KS)
potentials and to go beyond the popular random-phase approximation
for its functional derivative

The employed tdDFT scheme is summarised as follows:
\begin{enumerate}
\item Solve $\hat{h}\phi_i=\epsilon_i\phi_i$ for
\begin{align}
  \hat{h}\equiv&
  \frac{-1}{2}\nabla^2 + v_{\Ext}(\vr) + v_{\Hxc}^{\LEXX}[n](\vr)
  \label{eqn:hLEXX}
\end{align}
to determine the groundstate density $n=\sum_i F_i|\phi_i|^2$ and
Hartree, exchange and correlation (Hxc) potential
\begin{align}
  v_{\Hxc}^{\LEXX}[n](\vr)=\frac{\delta E_{\Hxc}^{\LEXX}[n]}{\delta n(\vr)}
  \label{eqn:vHxc}
\end{align}
using the LEXX approximation.
Here each orbital $i$ is assigned an occupation factor $F_i=2$
for the fully occupied inner orbitals and a value between
0 and 2 for the outermost orbital(s).%
\footnote{Generally only the frontier orbital is assigned
  a fractional occupation, but for certain transition metals both the
  outermost $s$ and $d$ shells are allowed to be partially occupied
  using the Hartree-Fock occupations.}
The Hartree, exchange and correlation energy is
\begin{align}
  E_{\Hxc}^{\LEXX}=&\int\frac{d\vr d\vrp}{2|\vr-\vrp|} n_{2\Hxc}^{\LEXX}(\vr,\vrp)
\end{align}
where the pair-density
$n_{2\Hxc}^{\LEXX}=\sum_{ij}F^S_{ij}n_i(\vr)n_j(\vrp)
  - F^U_{ij}\rho_i(\vr,\vrp)\rho_j(\vrp,\vr)$
is found via ensemble averaged
Hartree-Fock pair-densities for all degenerate states.
Here $\rho_i(\vr,\vrp)\equiv\phi^*_i(\vr)\phi_i(\vrp)$,
$n_i(\vr)\equiv\rho_i(\vr,\vr)$
and $F^S_{ij}$ and $F^U_{ij}$ are the ensemble-averaged
pair occupation factors for orbitals $i$ and $j$ which depend on
the degeneracy (including spin) and filling of the outermost orbital(s).
\item After using \eqref{eqn:hLEXX} to find the self-consistent
orbitals and density, solve
$[\hat{h}-\eta]G(\vr,\vrp;\eta)=\delta(\vr-\vrp)$
to obtain the response
\begin{align}
  \chi_0(\vr,\vrp;i\omega)=&2\Re\sum_i F_i\rho_i(\vr,\vrp)
  G(\vr,\vrp;\epsilon_i-i\omega)
  \\
  \equiv&
  -2\Re\sum_{ij}F_i\frac{\rho_i(\vr,\vrp)\rho_j(\vrp,\vr)}%
  {\epsilon_i-\epsilon_j-i\omega}
\end{align}
of the system to small changes in the Kohn-Sham potential
$v_s=v_{\Ext}+v_{\Hxc}$ at imaginary frequency $i\omega$.
\item Then solve\cite{RungeGross}
\begin{align}
  \chi =& \chi_0 + \chi_0\star f_{\Hxc}\star \chi
\end{align}
using $f\star g\equiv \int d\vr_1 f(\vr,\vr_1;\omega)g(\vr_1,\vrp;\omega)$
to find the response $\chi$ of the system to small changes in the
external potential $v_{\Ext}$. Here
\begin{align}
  f_{\Hxc}^{\LEXX}(\vr,\vrp;i\omega)\approx&
  \frac{\delta E_{\Hxc}^{\LEXX}}{\delta n(\vr)\delta n(\vrp)}|_{i\omega}
  \label{eqn:fHxc}
\end{align}
is the Hartree, exchange and correlation kernel associated with the
LEXX approximation.
\item Finally, after obtaining $\chi$, evaluate the imaginary frequency
dipole polarizabilities via
\begin{align}
  \alpha_Z(i\omega)=&\int d\vr d\vrp xx' \chi(\vr,\vrp;i\omega)
\end{align}
(for element $Z$ with $N=Z$ electrons)
and use the Casimir-Polder formula
\begin{align}
  C_{6ZZ'}=&\frac{3}{\pi}\int_0^{\infty}
  \alpha_Z(i\omega)\alpha_{Z'}(i\omega) d\omega
\end{align}
to determine the $C_6$ coefficients. Henceforth
we will use $\alpha_Z$ or $\alpha_Z(0)$ to mean the
$\omega=0$ polarizability
and $\alpha_Z(i\omega)$ to be the (imaginary) frequency
dependent polarizability.
\end{enumerate}

In the calculations carried out for this work
neither equations \eqref{eqn:vHxc} nor \eqref{eqn:fHxc}
are solved exactly.
In the former case the Krieger, Li and Iafrate\cite{KLI1992}
(KLI) approximation is employed while in the latter case
the Petersilka, Gossmann and Gross\cite{PGG} (PGG) kernel
\begin{align}
  f_{\Hxc}^{\text{PGG}}(\vr,\vrp)
  \approx&\frac{n_{2\Hxc}^{\LEXX}(\vr,\vrp)}{n(\vr)n(\vrp)}
\end{align}
or Radial Exchange-Hole\cite{Gould2012-RXH} (RXH) kernel
is used instead of the actual tdLEXX kernel.
The KLI approximation is expected
to make little difference to final results while the PGG and RXH
approximations will to contribute more substantially.
Nonetheless, both approximations avoid the worst
self-interaction and static correlation effects and
give generally good results\cite{Gould2016-C6}, comparable
to more sophisticated tdDFT
approaches\cite{Gould2012-2,Toulouse2013-RSH}.

Detailed technical details of all calculations are provided in
Refs.~\onlinecite{Gould2012-RXH,Gould2013-LEXX,%
Gould2013-Aff,Burke2016-Locality,Gould2016-C6}.
In short, all one and two-point quantities are expanded on
spherical harmonics, with the remaining radial functions
evaluated on grids. Unoccupied orbitals are avoided by
using shooting methods to evaluate Greens' functions. Numerical
errors are expected to be under 2\% (as a worst case),
even in the most polarizable of atoms.

\subsection{A special note on the transition metal atoms}

Because of the near-degeneracy of the $4/5s$ and $3/4d$ orbitals
in transition metals it is likely that some elements could
exhibit a discontinuous change\cite{Connerade2000}
in their Kohn-Sham orbital occupations as the confining potential
is varied.
To avoid these transitions, the occupation of the $4/5s$ and $3/4d$
orbitals was kept fixed throughout all confinements.
The occupation factors were generally kept in the
``groundstate'' arrangement i.e. the orbital arrangement that
gives the lowest energy in groundstate LEXX theory.
For Cr and Ag this caused problems for strong confinements
and the configurations [Ar]$4s^23d^4$ and [Kr]$5s^24d^0$
were used throughout the calculations.

\section{Results}

To study the behaviour of confined atoms, calculations were
carried out to test the effect of electron number and confining
potential on $V$, $\alpha$ and $C_6$. Simulations were performed
for atoms with one to 54 electrons, in confining potentials
$v_{\Ext}(\vr;r_c,k)=-N/r + r^{\k}/r_c^{\k+1}$
governed by $2\ls r_c\ls 30$ and $\k=2,3,4$.
Additional calculations were carried out with $r_c=\infty$
to reproduce the unconfined atom and provide a further check
on the confined calculations.
From the calculations the density and dipole polarizabilities
\begin{align}
  &n_Z(r;r_c,\k), &
  &\alpha_Z(i\omega;r_c,\k),
\end{align}
were found directly, while the volume
\begin{align}
  V_Z(r_c,\k)=&\int d\vr \frac{4\pi r^3}{3} n_Z(r;r_c,\k)
\end{align}
and same-species $C_6$ coefficients
\begin{align}
  C_{6ZZ}(r_c,\k)=&\frac{3}{\pi}\int_0^{\infty}\d\omega\alpha_Z(i\omega;r_c,\k)^2
\end{align}
were derived from the more basic ingredients.

\emph{It is worth highlighting an important point here}.
In the calculations carried out for this work the
non-outermost KS eigenvalues $\epsilon_i$
changed by up to $~10$~mHa with $r_c$.
This suggests that fixed core calculations could potentially be
problematic in studies of conventional, unconfined atoms.
Unfortunately, the all-electron code employed in these calculations
did not allow testing of this.

\subsection{Free atom properties}

\begin{figure}
  \includegraphics[width=\linewidth]{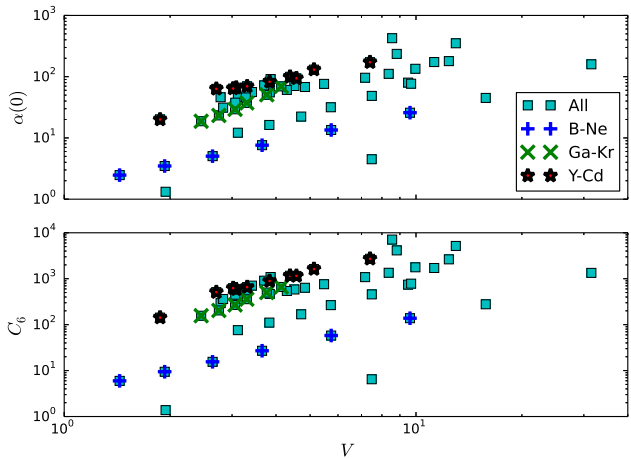}
  \caption{Log-log plots of polarizabilities $\alpha(0)$ (top)
    and $C_6$ coefficients (bottom) for
    atoms plotted as a function of volume. Selected open
    sub-shells are highlighted to show the effect of shell
    structure on the relationship between polarizability
    and volume.
    \label{fig:AtomsV}}
\end{figure}
As a first test let us study the properties of neutral atoms without
a confining potential; to test for
simple scaling laws related to atomic volume defined
via \eqref{eqn:V}. We plot in Figure~\ref{fig:AtomsV} the
bare polarizabilities $\alpha_Z$ and same-species vdW
coefficients $C_{6ZZ}$ as a function of $V_Z$ for all atoms
in the first five rows of the periodic tables i.e. $1\leq Z=N\leq 54$.

It is readily apparent that both $\alpha$ and $C_6$ are poorly
approximated by a single straight line, but could be
approximated by multiple straight lines in the plot with
different gradients  (corresponding to power laws with
different prefactors and exponents).
It is interesting to observe that the trends cluster
into groupings based on sub-shell structure (some examples are
highlighted in the plot). Clearly this behaviour is
unlikely to carry through into molecules. However, it does mean that
drawing conclusions from a limited subset of atoms is dangerous if one
does not take care to include a range of atomic types; and
that care needs to be taken when extrapolating atomic results
into atom-in-molecule calculations.

\subsection{Volume dependence}

Let us now begin to explore the effect of volume scaling
on individual atoms. By calculating
\begin{align}
  C_6(V;\k):=& C_6(V(r_c;\k);\k):= C_{6}(r_c,\k), \\
  \alpha(V;\k):=& \alpha(V(r_c;\k);\k):= \alpha(r_c,\k),
\end{align}
for selected values of $r_c$ the $C_6$ coefficients
may be evaluated as a function of volume.
The $C_6$ coefficients may then be fitted to the relationship
\begin{align}
  \frac{C_{6ZZ}(V_Z)}{C_{6ZZ}(V_Z^R)}
  \approx& \bigg( \frac{V_Z}{V_Z^R} \bigg)^{p_Z}
  \label{eqn:C6Fit}
\end{align}
by performing a linear fit of $\log(V/V^R)$ vs $\log(C_6/C_6^R)$.
Similarly the static polarizabilities can be fit to
\begin{align}
  \frac{\alpha_Z(V_Z)}{\alpha_Z(V_Z^R)}
  \approx& \bigg( \frac{V_Z}{V_Z^R} \bigg)^{p'_Z}.
  \label{eqn:A0Fit}
\end{align}

In the fits employed for this work, confined atoms with more
than a 50\% deviation
in volume from the unconfined $r_c\to\infty$
atom (i.e. for which $V(r_c)/V(\infty)< 0.5$) were discarded,
as these were often numerically unstable and are unlikely to be
a realistic representation of an atom in a molecules.
At least nine sampling points were included for all species,
and often more.
The power law fit is robust across all elements, giving
under 1.5\% root mean square errors for all but seven of the
elements, with a worst case for Rb ($N=37$) at 4.7\%.

\subsubsection{Variation with the confining exponent}

\begin{figure}
  \includegraphics[width=\linewidth]{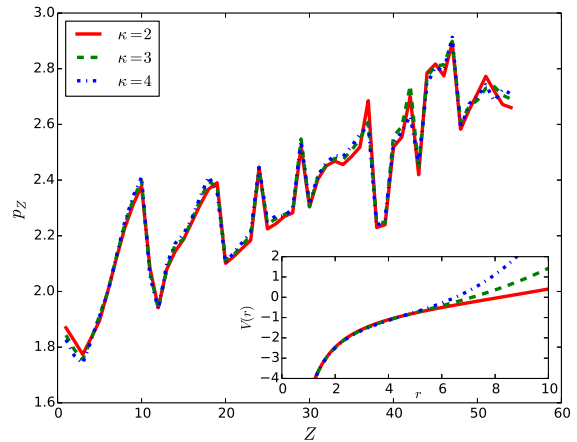}
  \caption{Exponents $p_Z$ for \eqref{eqn:C6Fit}
    as a function of $Z$, with $\k=2,3,4$.
    The invariance to the form of the confining potential
    suggests this is reasonable quantitative data.
    Inset shows the external potential for boron
    at 75\% volume and demonstrates that potentials with
    different $\k$ are substantially changed.
    \label{fig:CoeffsC6}}
\end{figure}
If the confining potential model is valid, it should give
consistent results as its exponent is changed.
The exponents $p_Z$ are shown in Figure~\ref{fig:CoeffsC6}
for the harmonic, cubic and quartic confining
potentials. It is clear from the plots that $p_Z$
is almost independent of the confining exponent $\k$,
suggesting that the form of the confining potential (at least
in the spherically symmetric ensemble case considered here) is largely
irrelevant. This is particularly surprising in the weakly confined
alkali metal atoms whose outermost electronst contribute
most or the polarizability and are highly diffuse.
In these cases one expects the contribution from
\begin{align}
  \left\langle \frac{r^{\k}}{r_c^{\k+1}}
  \right\rangle_h\equiv\int d\vr n_h(\vr)\frac{r^{\k}}{r_c^{\k+1}}
\end{align}
to be sensitive to $r_c$ and $\k$ separately.
%For illustration, the external potential that leads to boron
%having an effective volume 75\% of its unconfined volume is shown
%in the inset of Figure~\ref{fig:CoeffsC6}.

This insensitivity to the shape of the potential is very fortunate
for methods which employ a power law fit, even if they use a
species independent exponent.
It suggests that in the atom-in-molecule approximation, the behaviour
of the confined atom should be somewhat independent of the form
of the confinement, allowing the total polarizability of many
molecules to be written simply as a sum (or screened sum) of
scaled local polarizabilities.

\subsubsection{Variation with the kernel}

\begin{figure}[h!]
  \includegraphics[width=\linewidth]{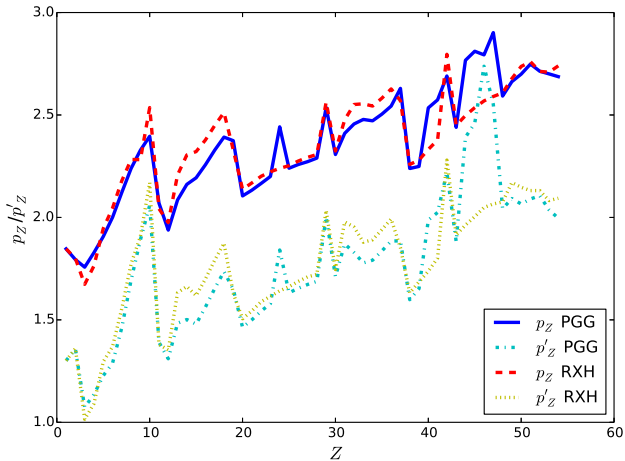}
  \caption{Scaling exponents $p_Z$ and $p'_Z$
    averaged over $\k=2,3,4$
    using the PGG\cite{PGG} and RXH\cite{Gould2012-RXH}
    kernel approximations. Results are largely insensitive to
    the kernel suggesting that the tdDFT results give
    reasonable quantitative data.\label{fig:RXH}
  }
\end{figure}
In a similar spirit to the previous test, let us now explore
the sensitivity of calculations to the tdDFT approximation.
If we are to expect qualitative accuracy from our data, it must
be largely independent of the kernel approximation chosen.
Thus, in addition to the calculations using the PGG approximation,
calculations using the radial-exchange hole (RXH)
kernel\cite{Gould2012-RXH} were carried out to test the sensitivity
of the static polarizabilities and $C_6$ coefficients to the kernel.

As can be seen in Figure~\ref{fig:RXH}, the scaling exponents are
not very sensitive to the type of kernel employed, although they are
more sensitive to the kernel than they are to the confining exponent.
The variation with kernel of the exponents is also small compared
to the variation with atomic number and shell structure.
This similarity appears despite
the different kernels producing substantially different polarizabilities
and $C_6$ coefficients
(see Tables~\ref{tab:Summary1}--\ref{tab:Summary3} in
Appendix~\ref{app:Summary}).

I speculate that the insensitivity to the kernel may be related to
the fact that the polarizability is dominated by the self-screened
outermost orbitals, which undergo similar (although certainly
not identical) dynamic screening in the RXH and PGG kernels i.e.
$n_h(\vr)n_h(\vrp)f_{\Hxc}^{\text{PGG}}(\vr,\vrp)
\approx n_h(\vr)n_h(\vrp)f_{\Hxc}^{\text{RXH}}(\vr,\vrp)$ where
$n_h$ is the electron density of the HOMO. This would also explain
why the greatest deviations between the two methods occur within
sub-shells where the difference between the detailed (PGG) and
radially averaged (RXH) pair densities are likely to be greatest.

This similarity may also responsible for the consistency of
the single-pole frequency
\begin{align}
  \Omega_Z=\frac{4C_{6ZZ}}{3\alpha_Z(0)^2},
  \label{eqn:Omega}
\end{align}
across the two approximations studied here and in Chu and Dalgarno's
results\cite{Chu2004}, as shown in
Tables~\ref{tab:Summary1}--\ref{tab:Summary3} in
Appendix~\ref{app:Summary}.
Here $\Omega$ is obtained from a [1,0]-Pad\'e approximation
$\alpha(i\omega)=\alpha(0)/(1+\omega^2/\Omega^2)$ to $\alpha(i\omega)$.
Regardless of its origins, the lack of sensitivity to the kernel
provides substantial reassurance that the results presented here
are not artefacts of the approximations employed.

\subsection{Behaviour of the scaling exponents}

\begin{figure}
  \includegraphics[width=\linewidth]{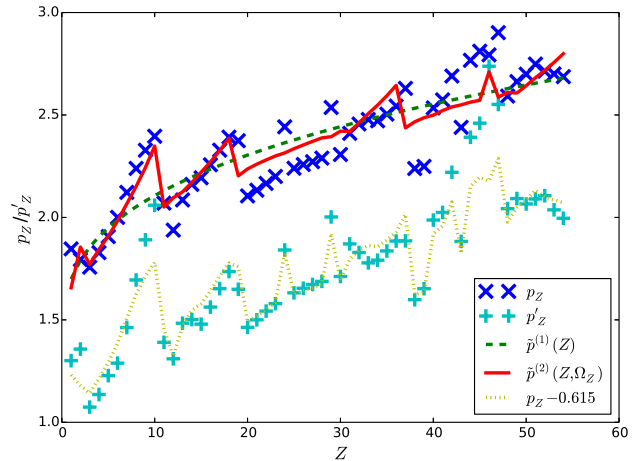}
  \caption{Exponent $p_Z$ versus $Z$,
    %plotted as a function of
    %$Z$ and $\Omega=\frac43 C_6/\alpha^2$ and
    compared to fits $\tilde{p}^{(1)}(Z)$ [equation~\eqref{eqn:p1}]
    taking into account variation with respect to electron number,
    and $\tilde{p}^{(2)}(Z,\Omega_Z)$ [equation~\eqref{eqn:p2}]
    also taking into account variation with the single-pole
    approximation coefficient $\Omega_Z$.
    Exponents $p'_Z$ for $\alpha$ are also plotted
    and compared with $p_Z-0.615$ to show the
    simple, and surprising, relationship between $p$ and $p'$.
    \label{fig:Fits}}
\end{figure}
Let us finally study the behaviour of the scaling exponents
themselves, to uncover key properties influencing their values.
Noting the sensitivity to shell structure of $\alpha$, $C_6$
and $V$ we propose a simple improvement to the single power law
$p=2$, by making $p$ depend on shell structure. We thus define
$\tilde{p}^{(0)}(n_Z)$ as the average coefficient over a given
shell. Unfortunately, this does not give terribly good results,
as shown in Table~\ref{tab:CoeffModel}.

\begin{table}
  \caption{Errors of different models of the scaling coefficient
    as a function of $Z$, and their mean absolute errors (MAEs).
    \label{tab:CoeffModel}}
  \begin{ruledtabular}%
    \footnotesize
    \begin{tabular}{rl|rrrr}
$Z$ & Sym. & $2-p_Z$ & $\tilde{p}^{(0)}-p_Z$ &
$\tilde{p}^{(1)}-p_Z$ & $\tilde{p}^{(2)}-p_Z$  \\
\hline
  1 &  H &  0.153 & -0.026 & -0.148 & -0.190  \\
  2 & He &  0.206 &  0.026 & -0.003 &  0.061  \\
\hline
  3 & Li &  0.243 &  0.315 &  0.098 &  0.013  \\
  4 & Be &  0.172 &  0.244 &  0.078 &  0.016  \\
  5 &  B &  0.093 &  0.166 &  0.043 &  0.002  \\
  6 &  C & -0.000 &  0.072 & -0.013 & -0.018  \\
  7 &  N & -0.122 & -0.050 & -0.101 & -0.055  \\
  8 &  O & -0.239 & -0.167 & -0.187 & -0.088  \\
  9 &  F & -0.328 & -0.256 & -0.248 & -0.084  \\
 10 & Ne & -0.396 & -0.324 & -0.290 & -0.048  \\
\hline
 11 & Na & -0.070 &  0.108 &  0.061 & -0.019  \\
 12 & Mg &  0.062 &  0.240 &  0.216 &  0.158  \\
 13 & Al & -0.085 &  0.093 &  0.091 &  0.040  \\
 14 & Si & -0.161 &  0.017 &  0.035 &  0.009  \\
 15 &  P & -0.193 & -0.015 &  0.023 &  0.027  \\
 16 &  S & -0.257 & -0.079 & -0.021 &  0.014  \\
 17 & Cl & -0.328 & -0.150 & -0.075 & -0.003  \\
 18 & Ar & -0.392 & -0.214 & -0.121 & -0.006  \\
\hline
 19 &  K & -0.374 & -0.031 & -0.087 & -0.172  \\
 20 & Ca & -0.105 &  0.239 &  0.199 &  0.132  \\
 21 & Sc & -0.133 &  0.211 &  0.187 &  0.127  \\
 22 & Ti & -0.166 &  0.178 &  0.169 &  0.115  \\
 23 &  V & -0.200 &  0.144 &  0.150 &  0.100  \\
 24 & Cr & -0.442 & -0.098 & -0.078 & -0.127  \\
 25 & Mn & -0.240 &  0.104 &  0.138 &  0.095  \\
 26 & Fe & -0.258 &  0.085 &  0.133 &  0.095  \\
 27 & Co & -0.272 &  0.072 &  0.133 &  0.099  \\
 28 & Ni & -0.290 &  0.054 &  0.128 &  0.099  \\
 29 & Cu & -0.537 & -0.193 & -0.106 & -0.143  \\
 30 & Zn & -0.307 &  0.037 &  0.136 &  0.114  \\
 31 & Ga & -0.410 & -0.067 &  0.044 &  0.009  \\
 32 & Ge & -0.456 & -0.112 &  0.011 &  0.005  \\
 33 & As & -0.478 & -0.134 &  0.000 &  0.028  \\
 34 & Se & -0.472 & -0.128 &  0.018 &  0.076  \\
 35 & Br & -0.504 & -0.160 & -0.003 &  0.090  \\
 36 & Kr & -0.543 & -0.200 & -0.032 &  0.101  \\
\hline
 37 & Rb & -0.630 &  0.005 & -0.108 & -0.193  \\
 38 & Sr & -0.239 &  0.396 &  0.294 &  0.228  \\
 39 &  Y & -0.249 &  0.386 &  0.294 &  0.238  \\
 40 & Zr & -0.535 &  0.100 &  0.019 & -0.036  \\
 41 & Nb & -0.574 &  0.061 & -0.011 & -0.052  \\
 42 & Mo & -0.690 & -0.055 & -0.117 & -0.151  \\
 43 & Tc & -0.440 &  0.196 &  0.143 &  0.110  \\
 44 & Ru & -0.767 & -0.132 & -0.175 & -0.204  \\
 45 & Rh & -0.811 & -0.176 & -0.210 & -0.240  \\
 46 & Pd & -0.793 & -0.158 & -0.183 & -0.083  \\
 47 & Ag & -0.902 & -0.267 & -0.282 & -0.315  \\
 48 & Cd & -0.592 &  0.043 &  0.037 &  0.018  \\
 49 & In & -0.663 & -0.028 & -0.025 & -0.056  \\
 50 & Sn & -0.699 & -0.064 & -0.053 & -0.056  \\
 51 & Sb & -0.748 & -0.113 & -0.093 & -0.067  \\
 52 & Te & -0.714 & -0.078 & -0.050 &  0.004  \\
 53 &  I & -0.701 & -0.065 & -0.029 &  0.056  \\
 54 & Xe & -0.687 & -0.052 & -0.007 &  0.112  \\
\hline
MAE &    &  0.391 &  0.133 &  0.106 &  0.089  \\
\hline
\end{tabular}
  \end{ruledtabular}
\end{table}
Going further, it is clear at a glance that the exponent grows with
$Z$ in a sub-linear fashion, and displays some of the
characteristics of the atomic sub-shells.
For the dependence on $Z$, a fit
\begin{align}
  \tilde{p}^{(1)}(Z)\approx& 1.346 + 0.353 Z^{1/3}
  \label{eqn:p1}
\end{align}
does a reasonable job of fitting the broad trend of the
exponents. However, as can be seen in Figure~\ref{fig:Fits}
and Table~\ref{tab:CoeffModel},
it misses the shell structure, leading to a
mean absolute error (MAE) of 0.106. Note that the dependency
on $Z^{1/3}$ was chosen based on a crude measure of atomic radius.

To improve this fit we note that $\Omega_Z$ [defined above
in \eqref{eqn:Omega}] is, like $p$ and $p'$, largely independent
of the kernel approximation employed. We thus introduce
a term depending on $\Omega_Z$. The resulting fit
\begin{align}
  \tilde{p}^{(2)}(Z,\Omega)\approx& 
  1.256 + [0.348 +0.122\Omega_Z] Z^{1/3}
  \label{eqn:p2}
\end{align}
is an improvement on \eqref{eqn:p1} with a MAE of 0.089
(see Table~\ref{tab:CoeffModel}).
While certainly imperfect, Figure~\ref{fig:Fits} shows
that $\tilde{p}^{(2)}(Z,\Omega_Z)$ is a decent approximation to $p_Z$,
introducing most of the ``spikes'' coming from shell structure.

A quick glance at Figure~\ref{fig:RXH} also highlights an additional
trend: the scaling exponent for the polarizability and $C_6$
coefficient appear to run together in parallel. Indeed
Figure~\ref{fig:Fits} makes it clear that the exponents are
approximately related by
\begin{align}
  p'_Z\approx& p_Z-0.615,
  \label{eqn:pp2}
\end{align}
giving $\tilde{p}^{(2)\prime}(Z,\Omega)\approx 0.641
+ [0.348 +0.122\Omega_Z] Z^{1/3}$
This result deviates from the expected $p'\approx p/2$ found by assuming
the scaling of $\alpha(i\omega)$ is the same for all $\omega$.
It follows that
\begin{align}
  \frac{C_{6ZZ}}{C_{6ZZ}^R}\approx&\frac{\alpha_Z}{\alpha_Z^R}
  \bigg( \frac{V_Z}{V_Z^R} \bigg)^{0.615}
\end{align}
is an almost-universal ($Z$ independent) relationship.

\section{Open questions}

Our numerical results point to complicated relationships
between $\alpha$, $C_6$ and atomic volume. Simple dimensional
arguments give rise to the incorrect $p_Z=2$ and $p_Z'=1$. Shell
structure effects are clearly visible in the coefficients.
However, even a relatively complex relationship \eqref{eqn:p2}
involving the three ``simple'' measures of polarizability is
insufficient to fully account for variations. 
The detailed mechanisms of polarizaztion under confinement thus
remains an open problem. They may be related
to the recently uncovered relationship
$C_6\propto \alpha(0)^{1.46}$ from \rcite{Gould2016-C6}.

On another front, this manuscript does not explore the more complex
case of spatial dependence in the confinement, and is not e.g.
able to explore differences between effective polarizabilities of
C atoms in diamond and graphite. These differences are especially
important in low-dimensional materials such as layers and nanotubes,
as first recognised in early studies
on layered geometries\cite{Dobson2006,Rydberg2003,Langreth2005}.
  
Although beyond the scope of the present work a similar analysis
with a non-symmetric confining potential would thus be very useful.
On a practical note it would uncover additional details of polarization
mechanisms. On a conceptual note it would explain whether embedded
atoms are dominated by Dobson-A (which are included in such a study)
or Dobson-B/Dobson-C (which are not) effects
(in the classification scheme of \rcite{Dobson2014-IJQC}),
and thus shed light on the role played by non-locality in
polarizable systems. Recent work\cite{Dobson2016-LRT}
allows direct comparisons to be made with higher-level theory.

\section{Conclusions}

In this work we have shown that $C_6$ coefficients of atoms in
subject to a confining potential designed to mimic the effect of a
molecule do scale as a power law of effective volume $V$. However,
the scaling exponents $p_Z$ depend on the number of electrons $Z$
and not with a species independent exponent of two as has previously
been assumed. We further showed that the exponent is almost
insensitive to the confining potentials tested, and only weakly
sensitive to the method used to evaluate $\alpha(i\omega)$, suggesting
that it this approximation is likely to hold for atoms in molecules,
at least up to ionic and other bonding effects.
%Aside from the data presented in
%Tables~\ref{tab:Summary1}-\ref{tab:Summary3},
Key new results are equations \eqref{eqn:p2} and \eqref{eqn:pp2}
showing the approximate relationship between the coefficients
$p_Z$ and $p'_Z$ and the number of electrons $N=Z$ and single-pole
frequency $\Omega=\frac43 C_6/\alpha^2$.

The work suggests that simple improvements might be made to
approximations based around the Becke-Johnson\cite{Becke2005-XDM},
Tkatchenko-Scheffler\cite{Tkatchenko2009},
or similar frameworks by simply modifiying the scaling
relationship. This might involve using the tabulated data
(Tables~\ref{tab:Summary1}--\ref{tab:Summary3})
or approximating the scaling exponent of a given atom by
Eq.~\eqref{eqn:p2} (which involves $Z$ and $\Omega_Z$).

The results presented here also raise
interesting questions about why polarizabilities behave the way
they do, and how that changes under confinement. For example
it would be interesting to understand why the outermost
orbitals and lowest excited states of confined atoms give rise
to features that behave largley independently of the form of
the confining potential and approximations made in their calculation.
This convenient property justifies the use of
time-dependent ensemble DFT calculations in the linear-response
regime for studying
atoms in confined potentials. Especially
when combined with more accurate
(but more difficult) calculations of unconfined atoms.

\acknowledgments

The author would like to thank R.~A.~DiStasio~Jr,
Erin Johnson, and T.~C.~V.~Bu\v{c}ko for helpful discussion.
TG received computing support from the Griffith University
Gowonda HPC Cluster. 

\appendix

\section{Data summary}
\label{app:Summary}

In Tables~\ref{tab:Summary1}-\ref{tab:Summary3} we show static
polarizabilites, $C_6$ coefficients and single-pole frequency
$\Omega_Z=4C_{6ZZ}/[3\alpha_Z(0)^2]$
from the two approximations studied in the manuscript, and from
Chu and Dalgarno\cite{Chu2004}. It is worth noting the robustness
of $\Omega_Z$ across the three approximations, compared to $\alpha$
and $C_6$ which vary considerably.

\begin{table}[ht]
  \caption{Summary data for Rows 1-3.
    Benchmark data (BM) is from the ``Recommended'' and ``Corrected''
    results of Chu and Dalgarno\cite{Chu2004},
    for $\alpha(0)$ and $C_6$ respectively, and is provided for
    comparison.
    Units are: $C_6$ [Ha$a_0^6$], $\alpha(0)$ [$a_0^3$],
    $\Omega$ [Ha], $p$ [unitless].
    \label{tab:Summary1}
  }
  \begin{ruledtabular}%
    \footnotesize
    \begin{tabular}{l|rrr|r|rrr|r|r}
 & \multicolumn{4}{c|}{PGG}
 & \multicolumn{4}{c|}{RXH}
 & BM \\
\hline
$Z$ & $C_6$ & $\alpha(0)$ & $\Omega$ & $p$
 & $C_6$ & $\alpha(0)$ & $\Omega$ & $p$
 & $\Omega$ \\
\hline
 H &      6.5 &      4.5 &  0.43 &  1.85 &      6.5 &      4.5 &  0.43 &  1.85 &  0.43\\ 
He &      1.4 &      1.3 &  1.05 &  1.79 &      1.4 &      1.3 &  1.04 &  1.80 &  1.02\\ 
\hline
Li &   1341.0 &    159.2 &  0.07 &  1.76 &   1330.9 &    158.7 &  0.07 &  1.67 &  0.07\\ 
Be &    277.4 &     45.0 &  0.18 &  1.83 &    249.3 &     41.9 &  0.19 &  1.76 &  0.20\\ 
\hline
 B &    138.0 &     25.8 &  0.28 &  1.91 &    100.7 &     21.0 &  0.31 &  1.95 &  0.30\\ 
 C &     58.0 &     13.5 &  0.43 &  2.00 &     43.8 &     11.2 &  0.47 &  2.05 &  0.43\\ 
 N &     27.0 &      7.6 &  0.62 &  2.12 &     21.4 &      6.5 &  0.67 &  2.18 &  0.59\\ 
 O &     15.5 &      5.0 &  0.82 &  2.24 &     13.0 &      4.5 &  0.85 &  2.28 &  0.71\\ 
 F &      9.4 &      3.5 &  1.04 &  2.33 &      7.6 &      3.1 &  1.08 &  2.28 &  0.88\\ 
Ne &      6.0 &      2.5 &  1.30 &  2.40 &      5.3 &      2.3 &  1.33 &  2.54 &  1.20\\ 
\hline
Na &   1711.6 &    173.3 &  0.08 &  2.07 &   1187.7 &    135.9 &  0.09 &  2.04 &  0.08\\ 
Mg &    743.5 &     80.1 &  0.15 &  1.94 &    519.0 &     63.3 &  0.17 &  1.97 &  0.16\\ 
\hline
Al &    781.0 &     76.9 &  0.18 &  2.09 &    494.5 &     58.3 &  0.19 &  2.21 &  0.20\\ 
Si &    455.8 &     48.6 &  0.26 &  2.16 &    303.0 &     37.8 &  0.28 &  2.31 &  0.30\\ 
 P &    265.8 &     31.6 &  0.35 &  2.19 &    186.0 &     25.3 &  0.39 &  2.32 &  0.39\\ 
 S &    167.9 &     22.3 &  0.45 &  2.26 &    119.5 &     18.1 &  0.49 &  2.38 &  0.47\\ 
Cl &    111.1 &     16.3 &  0.56 &  2.33 &     80.8 &     13.4 &  0.60 &  2.45 &  0.58\\ 
Ar &     75.9 &     12.2 &  0.68 &  2.39 &     56.3 &     10.1 &  0.73 &  2.51 &  0.70\\ 
\end{tabular}

  \end{ruledtabular}
\end{table}
\begin{table}[ht]
  \caption{Same as table~\ref{tab:Summary1} but for Row 4.
    \label{tab:Summary2}
  }
  \begin{ruledtabular}%
    \footnotesize
    \begin{tabular}{l|rrr|r|rrr|r|r}
 & \multicolumn{4}{c|}{PGG}
 & \multicolumn{4}{c|}{RXH}
 & BM \\
\hline
$Z$ & $C_6$ & $\alpha(0)$ & $\Omega$ & $p$
 & $C_6$ & $\alpha(0)$ & $\Omega$ & $p$
 & $\Omega$ \\
\hline
 K &   5172.2 &    351.0 &  0.06 &  2.37 &   3437.8 &    267.1 &  0.06 &  2.33 &  0.06\\ 
Ca &   2637.5 &    179.4 &  0.11 &  2.11 &   1764.9 &    137.6 &  0.12 &  2.14 &  0.11\\ 
\hline
Sc &   1771.9 &    134.6 &  0.13 &  2.13 &   1234.7 &    106.0 &  0.15 &  2.17 &  0.13\\ 
Ti &   1348.3 &    111.4 &  0.14 &  2.17 &    965.3 &     89.4 &  0.16 &  2.20 &  0.14\\ 
 V &   1082.0 &     96.2 &  0.16 &  2.20 &    789.5 &     78.2 &  0.17 &  2.22 &  0.16\\ 
Cr &    582.7 &     70.4 &  0.16 &  2.44 &    663.1 &     69.8 &  0.18 &  2.24 &  0.13\\ 
Mn &    758.1 &     76.3 &  0.17 &  2.24 &    567.0 &     63.1 &  0.19 &  2.25 &  0.19\\ 
Fe &    635.7 &     67.8 &  0.18 &  2.26 &    478.2 &     56.3 &  0.20 &  2.28 &  0.20\\ 
Co &    540.7 &     60.9 &  0.19 &  2.27 &    409.0 &     50.8 &  0.21 &  2.29 &  0.22\\ 
Ni &    466.5 &     55.3 &  0.20 &  2.29 &    354.9 &     46.2 &  0.22 &  2.31 &  0.22\\ 
Cu &    293.0 &     46.1 &  0.18 &  2.54 &    245.2 &     41.2 &  0.19 &  2.56 &  0.19\\ 
Zn &    357.9 &     46.5 &  0.22 &  2.31 &    275.2 &     39.2 &  0.24 &  2.32 &  0.24\\ 
\hline
Ga &    661.3 &     69.4 &  0.18 &  2.41 &    461.6 &     56.0 &  0.20 &  2.48 &  0.18\\ 
Ge &    500.2 &     50.7 &  0.26 &  2.46 &    355.0 &     41.2 &  0.28 &  2.55 &  0.28\\ 
As &    358.3 &     37.3 &  0.34 &  2.48 &    261.2 &     30.7 &  0.37 &  2.55 &  0.39\\ 
Se &    268.7 &     29.3 &  0.42 &  2.47 &    194.2 &     24.0 &  0.45 &  2.54 &  0.45\\ 
Br &    203.8 &     23.3 &  0.50 &  2.50 &    148.2 &     19.2 &  0.54 &  2.58 &  0.54\\ 
Kr &    156.5 &     18.7 &  0.59 &  2.54 &    115.0 &     15.5 &  0.64 &  2.63 &  0.61\\ 
\end{tabular}

  \end{ruledtabular}
\end{table}
\begin{table}[ht]
  \caption{Same as table~\ref{tab:Summary1} but for Row 5.
    \label{tab:Summary3}
  }
  \begin{ruledtabular}%
    \footnotesize
    \begin{tabular}{l|rrr|r|rrr|r|r}
 & \multicolumn{4}{c|}{PGG}
 & \multicolumn{4}{c|}{RXH}
 & BM \\
\hline
$Z$ & $C_6$ & $\alpha(0)$ & $\Omega$ & $p$
 & $C_6$ & $\alpha(0)$ & $\Omega$ & $p$
 & $\Omega$ \\
\hline
Rb &   7118.2 &    424.7 &  0.05 &  2.63 &   4754.3 &    324.3 &  0.06 &  2.57 &  0.06\\ 
Sr &   4132.0 &    235.7 &  0.10 &  2.24 &   2750.1 &    180.1 &  0.11 &  2.26 &  0.11\\ 
\hline
 Y &   2714.2 &    172.2 &  0.12 &  2.25 &   1889.4 &    134.9 &  0.14 &  2.28 &    --\\ 
Zr &   1624.2 &    130.5 &  0.13 &  2.53 &   1454.1 &    109.8 &  0.16 &  2.33 &    --\\ 
Nb &   1159.4 &     99.6 &  0.16 &  2.57 &   1189.8 &     95.5 &  0.17 &  2.38 &    --\\ 
Mo &    888.7 &     82.7 &  0.17 &  2.69 &    825.9 &     79.0 &  0.18 &  2.80 &    --\\ 
Tc &   1150.0 &     93.9 &  0.17 &  2.44 &    876.7 &     78.5 &  0.19 &  2.45 &    --\\ 
Ru &    656.1 &     69.5 &  0.18 &  2.77 &    754.7 &     71.4 &  0.20 &  2.50 &    --\\ 
Rh &    589.6 &     66.2 &  0.18 &  2.81 &    661.6 &     65.7 &  0.20 &  2.54 &    --\\ 
Pd &    142.5 &     20.0 &  0.47 &  2.79 &    588.1 &     61.1 &  0.21 &  2.57 &    --\\ 
Ag &    509.7 &     63.3 &  0.17 &  2.90 &    528.8 &     57.3 &  0.22 &  2.59 &  0.21\\ 
Cd &    623.4 &     64.1 &  0.20 &  2.59 &    480.0 &     54.0 &  0.22 &  2.61 &    --\\ 
\hline
In &   1088.8 &     91.5 &  0.17 &  2.66 &    755.3 &     73.1 &  0.19 &  2.68 &  0.18\\ 
Sn &    908.0 &     72.0 &  0.23 &  2.70 &    637.7 &     57.9 &  0.25 &  2.74 &  0.24\\ 
Sb &    707.7 &     56.2 &  0.30 &  2.75 &    508.8 &     45.7 &  0.33 &  2.76 &  0.34\\ 
Te &    560.8 &     45.8 &  0.36 &  2.71 &    398.6 &     36.9 &  0.39 &  2.71 &  0.37\\ 
 I &    448.6 &     37.6 &  0.42 &  2.70 &    320.3 &     30.5 &  0.46 &  2.71 &  0.45\\ 
Xe &    362.2 &     31.3 &  0.49 &  2.69 &    260.9 &     25.4 &  0.54 &  2.74 &  0.52\\ 
\end{tabular}

  \end{ruledtabular}
\end{table}

\section*{References}

\end{document}